\documentclass[aps,prd,twocolumn,groupedaddress]{revtex4-1}

\usepackage{amssymb}
\usepackage{graphicx}
\usepackage{dcolumn}
\usepackage{bm}
\usepackage{hyperref}

\begin{document}

\title{Observational Tests for Oscillating Expansion Rate of the Universe}

\author{Koichi Hirano}
\email[]{Electronic address: hirano@ichinoseki.ac.jp}
\affiliation{Department of Physics, Ichinoseki National College of Technology, Ichinoseki 021-8511, Japan}

\author{Zen Komiya}
\affiliation{Department of Physics, Tokyo University of Science, Tokyo 162-8601, Japan}

\date{\today}

\begin{abstract}
We investigate the observational constraints on the oscillating scalar field model using data from type Ia supernovae, cosmic microwave background anisotropies, and baryon acoustic oscillations. According to a Fourier analysis, the galaxy number count $N$ from redshift $z$ data indicates that galaxies have preferred periodic redshift spacings. We fix the mass of the scalar field as $m_\phi=3.2\times 10^{-31}h$ ${\rm eV}$ such that the scalar field model can account for the redshift spacings, and we constrain the other basic parameters by comparing the model with accurate observational data. We obtain the following constraints: $\Omega_{m,0}=0.28\pm 0.03$ (95\% C.L.), $\Omega_{\phi,0} < 0.035$ (95\% C.L.), $\xi > -158$ (95\% C.L.) (in the range $\xi \le 0$). The best fit values of the energy density parameter of the scalar field and the coupling constant are $\Omega_{\phi,0}= 0.01$ and $\xi= -25$, respectively. The value of $\Omega_{\phi,0}$ is close to but not equal to $0$. Hence, in the scalar field model, the amplitude of the galaxy number count cannot be large. However, because the best fit values of $\Omega_{\phi,0}$ and $\xi$ are not $0$, the scalar field model has the possibility of accounting for the periodic structure in the $N$--$z$ relation of galaxies. The variation of the effective gravitational constant in the scalar field model is not inconsistent with the bound from observation.
\end{abstract}


\maketitle

\section{INTRODUCTION}

A widespread idea in cosmology is that the universe is homogeneous and isotropic above a certain scale. This hypothesis, usually called the cosmological principle (e.g., \cite{pee1993}), is thought to be a generalization of the Copernican principle that ``the Earth is not in a central, specially favored position''. The assumption is that any observer at any place at the same epoch would see essentially the same picture of the large scale distribution of galaxies in the universe.

However, according to a Fourier analysis by Hartnett \& Hirano \cite{har2008}, the galaxy number count $N$ from redshift $z$ data ($N$--$z$ relation) indicates that galaxies have preferred periodic redshift spacings of $\Delta z$ = 0.0102, 0.0246, and 0.0448 in the Sloan Digital Sky Survey (SDSS) (see Fig. 2 of \cite{teg2004}), with very similar results from the 2dF Galaxy Redshift Survey (2dF GRS) (see Fig. 17 of \cite{col2001}) \cite{har2008}. These redshift spacings have been confirmed by mass density fluctuations, the power spectrum $P(z)$, and $N_{pairs}$ calculations \cite{har2008}. The combined results from both surveys give characteristic periods of $31.7 \pm 1.8~h^{-1}~{\rm Mpc}$, $73.4 \pm 5.8~h^{-1}~{\rm Mpc}$, and $127 \pm 21~h^{-1}~{\rm Mpc}$ \cite{har2008}. That is, the redshift space for relatively high galaxy number count and other that exhibits comparatively low number counts appear alternately.

$127~h^{-1}~{\rm Mpc}$ is the same scale as that found in a pencil-beam survey of field galaxies \cite{bro1990}. Furthermore, the periodicity as a function of $z$ in the distribution of QSO spectra has also been reported \cite{har2009, rya2010}.

A natural interpretation is that concentric spherical shells of higher galaxy number densities surround us, with their individual centers situated at our location. However, if this interpretation reflected the actual physical concentration of galaxies existing at certain distances from us, it would definitely be incompatible with the cosmological principle that presumes uniformity and isotropy of our space--time. In fact, it has been demonstrated \cite{yos2001}, from many numerical simulations using the Einstein--de Sitter and $\Lambda$CDM models, that the probability of getting such a periodic spatial structure from clustering and cosmic web filaments is less than $10^{-3}$.

One viable explanation for the periodicity of redshift space (the picket-fence structure of the $N$--$z$ relation) that preserves the cosmological principle would be introducing a scalar field that non-minimally couples with the curvature scalar, as originally proposed by Morikawa (1990 ,1991) \cite{mor1990, mor1991} and extended later by Hirano et al. (2008) \cite{hir2008a}. Such coupling can produce a periodic structure of redshift space in the following way. These scalar field models \cite{mor1990, mor1991, hir2008a} assume a potential $V(\phi)$ proportional to $\phi^2$ so that the scalar field oscillates around the potential minimum, inducing an oscillation of the Hubble parameter, due to the curvature coupling of $\phi$. An epoch of relatively rapid expansion and an epoch of relatively slow expansion appear alternately due to this oscillation. During the former epoch, the number density of galaxies diminishes noticeably, whereas during the latter epoch, the decrease in the number density is less drastic. This would then give a periodic structure as an apparent or illusionary effect in the $N$--$z$ relation, because we are observing the two types of epoch alternately with increasing $z$. Since the spatial distribution of galaxies remains homogeneous for a given instance of time, the cosmological principle can still be preserved.

In scalar field models \cite{mor1990, mor1991, hir2008a}, the scale of the periodic redshift spacings is determined by the mass parameter $m_\phi$ of the scalar field. In this paper, we first fix the mass of the scalar field as $m_\phi=3.2\times 10^{-31}h~{\rm eV}$ such that the scalar field model can account for the redshift spacings of the $127 h^{-1}~{\rm Mpc}$ period \cite{har2008}. Next, we constrain the other basic parameters of the scalar field model of Hirano et al. (2008) \cite{hir2008a} by comparing the model with accurate observational data from type Ia supernovae (SN Ia) \cite{hic2009}, cosmic microwave background (CMB) anisotropies \cite{kom2009}, and baryon acoustic oscillations (BAO) \cite{eis2005}. Finally, we test the validity of this model on variation of the effective gravitational constant.

This paper is organized as follows. In the next section we summarize the evolution equations for the scalar field and the scale factor of the oscillating scalar field model of Hirano et al. (2008) \cite{hir2008a} and describe the stationary state of the scalar field. In Section \ref{obstest} we investigate the observational constraints on the scalar field model \cite{hir2008a} using data from SN Ia \cite{hic2009}, CMB anisotropies \cite{kom2009}, and BAO \cite{eis2005}. For this model, we investigate variation of the effective gravitational constant in order to test the validity of the model. Finally, the conclusions are given in Section \ref{conclu}.

\section{OSCILLATING SCALAR FIELD MODEL}

\subsection{Basic Equations}
The action $S$ for our cosmological model is given by \cite{hir2008a}:
\begin{eqnarray}
S & = & \int d^4 x \sqrt{-g}\ \left[\frac{1}{2}\xi R\psi^2-\frac{c^4}{16\pi G}R+\frac{1}{2}g^{\mu\nu}\partial_{\mu}\psi\partial_{\nu}\psi\right. \nonumber \\
  &   & -\frac{1}{2}\left(\frac{m_\phi c}{\hbar}\right)^2\psi^2\exp{\left(-q\frac{4\pi G}{3c^4}\psi^2\right)} \nonumber \\
  &   & \left. +\frac{c^4}{8\pi G}\mit\Lambda +L^{(\rm mr)}\right], \label{Lagrangean}
\end{eqnarray}
where $\psi$ is the scalar field, $R$ is the scalar curvature, $\xi$ is the coupling constant, $m_\phi$ is the mass of the scalar field, $q$ is a scalar constant, $g^{\mu\nu}$ are the $(\mu,\nu)$-components of the metric tensor, $G$ is Newton's gravitational constant, $c$ is the velocity of light, $\mit\Lambda$ is Einstein's cosmological constant, and $L^{(\rm mr)}$ is the Lagrangian due to matter and radiation.
  
Based on the principle of least action, we extremize the action integral by setting $\delta S=0$, which gives the following gravitational field equations: 
\begin{eqnarray}
\lefteqn{R_{\mu\nu}-\frac{1}{2}g_{\mu\nu}R+\mit\Lambda g_{\mu\nu}} \nonumber \\
& = &\frac{8\pi G}{c^4}\Bigg[\partial_{\mu}\psi\partial_{\nu}\psi-\frac{1}{2}g_{\mu\nu}\partial_{\rho}\psi\partial^{\rho}\psi  \Bigg. \nonumber \\
&   &+\xi g_{\mu\nu}\Box\psi^2-\xi[\psi^2]_{;\mu\nu}+\xi\psi^2\left(R_{\mu\nu}-\frac{1}{2}g_{\mu\nu}R\right) \nonumber \\
&   &+\frac{1}{2}g_{\mu\nu}\left(\frac{m_\phi c}{\hbar}\right)^2 \psi^2\exp{\left(-q\frac{4\pi G}{3c^4}\psi^2\right)} \nonumber \\
&   & \Bigg.+T^{(\rm mr)}_{\mu\nu} \Bigg], \label{graveq}
\end{eqnarray}
where the symbol~``$\Box$'' signifies the d'Alembertian, the symbol~``$(\cdots)_{;\mu}$'' is the covariant derivative of the quantity ``$\cdots$'' with respect to the spatial coordinate $x^\mu$, and the quantities 
$T^{(\rm mr)}_{\mu\nu}$ are the $(\mu,\nu)$-components of the energy--momentum tensor of  matter and radiation.

Note that the potential term (the fourth term on the right-hand side of Eq. (\ref{Lagrangean})) reflects our modification \cite{hir2008a} to the original form employed by Morikawa (1990, 1991) \cite{mor1990,mor1991}, which coincides with the latter if $q=0$. This new form of the potential enables us to control the epoch when the growth of the scalar field $\psi$ begins to take place. The closer this epoch is to the present era, the less affected are the amplitudes of the spatial power spectrum of the CMB temperature anisotropy in the large-scale domain because of the reduced effect of the late-time integrated Sachs--Wolfe effect \cite{hir2008a}.
In the observational constraints of this paper, because we want to know the value of the basic parameters of the scalar field model, we fix the constant parameter as $q=0$.

By varying the action (Eq. (\ref{Lagrangean})) with respect to $\psi$, we obtain the time evolution equation for the scalar field:
\begin{equation}
\ddot{\phi}=-2\frac{\dot{a}}{a}\dot{\phi}-6\xi\frac{\ddot{a}}{a}\phi-a^2\left(\frac{m_\phi c}{\hbar}\right)^2(1-q\phi^2)\phi\exp{(-q\phi^2)}, \label{EqPhidots}
\end{equation}
where we have employed a dimensionless quantity $\phi$ instead of $\psi$:
\begin{equation}
\displaystyle{ \phi=\sqrt{{4\pi G\over 3c^4}}~\psi}. \label{eq:psiphi}
 \end{equation}
We also employ the short-hand notation $\dot{A}\equiv {\rm d}A/{\rm d}t_{\rm c}$,
where $t_{\rm c}$ is the conformal time defined as
\begin{equation}
 t_{\rm c} = \int\frac{c}{a}{\rm d}t,~~~~{\rm so~that}~~ 
\dot{A}=\frac{{\rm d}A}{{\rm d}t_c}=\frac{a}{c}\frac{{\rm d}A}{{\rm d}t},
\end{equation}
where $t$ is the proper time.

Assuming the Robertson--Walker line element, from the (0,0)-components and (1,1)-components of Eq. (\ref{graveq}), and using Eq. (\ref{EqPhidots}), we obtain the following set of time evolution equations for the cosmic scale factor $a$:
\begin{eqnarray}
\frac{\dot{a}}{a} & = & \Bigg[6\xi\phi\dot{\phi}+\left\{(6\xi\phi\dot{\phi})^2+(1-6\xi\phi^2) \  \right. \nonumber \\
                  &   & \times\left((\dot{\phi})^2+a^2\left(\frac{m_\phi c}{\hbar}\right)^2\phi^2\exp(-q\phi^2) \right.\nonumber \\
                  &   & \left.\left.+\frac{H_0^2}{c^2}\left(\frac{\Omega_{m,0}}{a}+\frac{\Omega_{r,0}}{a^2}+\Omega_{{\mit\Lambda},0}a^2\right)\right)\right\}^{1/2}\Bigg] \nonumber \\
                  &   & / (1-6\xi\phi^2),     \label{EqAdot}
\end{eqnarray}
\begin{eqnarray}
\frac{\ddot{a}}{a} & = & \Bigg[2\frac{\dot{a}^2}{a^2}(1-6\xi\phi^2)-3(\dot{\phi})^2-24\xi\frac{\dot{a}}{a}\phi\dot{\phi}+6\xi(\dot{\phi})^2 \nonumber \\
                   &   & -6\xi a^2\left(\frac{m_\phi c}{\hbar}\right)^2(1-q\phi^2)\phi^2\exp{(-q\phi^2)} \nonumber \\
                   &   & -\frac{3}{2}\frac{H_0^2}{c^2}\left(\frac{\Omega_{m,0}}{a}+\frac{\Omega_{r,0}}{a^2}\right)\Bigg] \nonumber \\
                   &   & / \{1-6\xi\phi^2(1-6\xi)\},   \label{EqAdots}
\end{eqnarray}
where $H_0$ is the Hubble constant. The quantities $\Omega_{\rm m,0}$, $\Omega_{\rm r,0}$, and $\Omega_{{\mit\Lambda},0}$ are the present mass densities of matter, radiation, and the cosmological constant $\mit\Lambda$ normalized, respectively, to the present value of the critical density $\rho_{\rm c,0}(=3H_0^2/8\pi G)$, viz., $\Omega_{\rm m,0}\equiv\rho_{\rm m,0}/\rho_{\rm c,0}$, $\Omega_{\rm r,0}\equiv\rho_{\rm r,0}/\rho_{\rm c,0}$ and $\Omega_{{\mit\Lambda},0}=(c/H_0)^2{\mit\Lambda}/3$. The matter included in our model universe is assumed to consist of baryonic matter as well as cold dark matter (CDM). 

The corresponding density $\rho_\phi$ and the pressure $P_\phi$ for the scalar field $\phi$ are given by ${\rm diag}T^\nu_\mu = (\rho, p, p, p)$ with 
\begin{eqnarray}
\rho_{\phi}c^2  =  \frac{3c^4}{4\pi G} & & \left[\frac{1}{2}\frac{1}{a^2}(\dot{\phi})^2+3\xi\frac{(\dot{a})^2}{a^4}\phi^2+6\xi\frac{\dot{a}}{a^3}\phi\dot{\phi}\right. \nonumber \\
               &   & \left.+\frac{1}{2}\left(\frac{m_\phi c}{\hbar}\right)^2\phi^2\exp{\left(-q\phi^2\right)}\right], \label{rhopsi}
\end{eqnarray}
\begin{eqnarray}
P_{\phi} = \frac{3c^4}{4\pi G} & & \left[\frac{1}{2}\frac{1}{a^2}(\dot{\phi})^2-2\xi\frac{\ddot{a}}{a^3}\phi^2+\xi\frac{(\dot{a})^2}{a^4}\phi^2\right.  \nonumber \\
    & & -2\xi\frac{1}{a^2}\phi\ddot{\phi}-2\xi\frac{1}{a^2}(\dot{\phi})^2-2\xi\frac{\dot{a}}{a^3}\phi\dot{\phi} \nonumber \\
    & & \left.-\frac{1}{2}\left(\frac{m_\phi c}{\hbar}\right)^2\phi^2\exp{\left(-q\phi^2\right)}\right].  \label{Ppsi}
\end{eqnarray}

From Eqs. (\ref{rhopsi}) and (\ref{Ppsi}) the scalar field satisfies the continuity equation:
\begin{eqnarray}
\dot{\rho_{\phi}}c^2&+&3\frac{\dot{a}}{a}(\rho_{\phi} c^2+P_{\phi})=0. \label{eqstatepsi}
\end{eqnarray}
There is no mutual energy exchange between the matter, radiation, and scalar fields. The scalar field interacts with gravity merely by coupling with the scalar curvature. 

The density parameter of the scalar field $\phi$ for the present epoch is defined as $\Omega_{\phi,0}\equiv\rho_{\phi,0}/\rho_{\rm c,0}$. Furthermore, since we are concerned only with a flat geometry, we obtain the following constraint from Eq. (\ref{EqAdot}) for the present era:
\begin{equation}
\Omega_{m,0}+\Omega_{r,0}+\Omega_{\phi,0}+\Omega_{{\mit\Lambda},0} = 1.
\end{equation}

Measurements of distant type Ia supernovae (SN Ia) \cite{rie1998, per1999} indicate late-time accelerated expansion of the universe. For the mechanism for the acceleration, there are currently many models including dark energy \cite{kom2006, kom2005} or modified gravity \cite{hir2010a, hir2010b}. In our scalar field model, it is always the cosmological constant which supplies nearly all of the driving force to the acceleration.

\subsection{Numerical Computation}

Salgado et al. (1996) \cite{sal1996} and Quevedo et al. (1997) \cite{que1997} found that the magnitude of the scalar field must remain extremely small during the epochs around the Big Bang nucleosynthesis in order not to effect the successful prediction made for the cosmic abundances of light elements based on the Einstein--de Sitter model universe. Further, according to Hirano et al. (2006) \cite{hir2006}, unless the value of $\phi$ is kept stationary at almost $0$ for a sufficiently long time in the early stage of expansion, the WMAP observation of the CMB angular spectrum cannot be reproduced because of the enhanced effect of the late-time integrated Sachs--Wolfe effect. Therefore, in what follows, we consider only models having a stationary state of the scalar field $\phi$ in the early epoch of the universe. In the case of having a stationary state of the scalar field, because the value of the scalar field $\phi$ is kept stationary at almost $0$ in the early epoch of the universe, the effective gravitational constant $G_{\rm eff}$ in the era of the big bang nucleosynthesis is almost equal to the value of the Newton's gravitational constant $G$.

To create such a stationary state of $\phi$ at those periods of time without any fine-tuning of the present-day value of $\phi_0$ such as those carried out by Salgado et al. (1996) \cite{sal1996}, we integrate the evolution equations in the direction from past to present (natural direction).
 
\begin{figure}[h!]
\includegraphics[width=90mm]{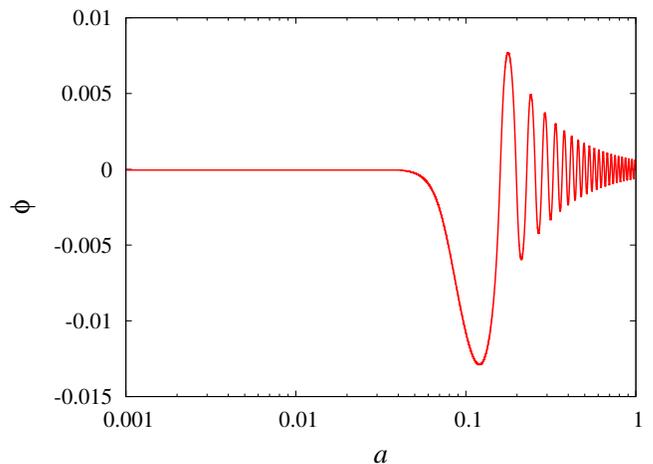}
\caption{An example of the variation of the scalar field $\phi$ as a function of scale factor $a$, starting from a stationary state. The parameters are $\Omega_{\rm m,0}=0.28$, $\Omega_{\phi,0}=0.01$, $\xi=-25$, $m_\phi=3.2\times 10^{-31}h$~eV, $q=0$, and $H_0=72$ ${\rm km~s^{-1} Mpc^{-1}}$.
\label{fig:phi}}
\end{figure}

A sample behavior of $\phi$ for the oscillating scalar field model of Hirano et al. (2008) \cite{hir2008a} is shown in Fig. \ref{fig:phi} as a function of scale factor $a$. We have used $\Omega_{\rm m,0}=0.28$, $\Omega_{\phi,0}=0.01$, $\xi=-25$, $m_\phi=3.2\times 10^{-31}h$~eV, $q=0$, and $H_0=72$ ${\rm km~s^{-1} Mpc^{-1}}$.

\begin{figure}[h!]
\includegraphics[width=90mm]{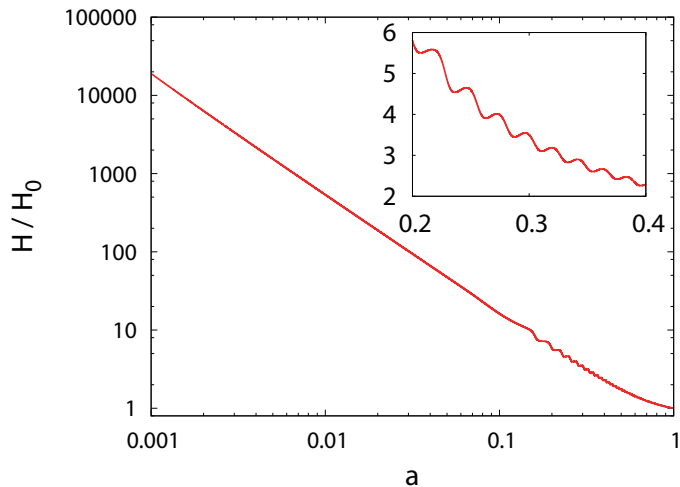}
\caption{Oscillation of the Hubble parameter normalized to the Hubble constant $H/H_0$. Top right panel is a close-up view. The abscissa is the scale factor $a$. The parameters are the same as these of Fig. \ref{fig:phi}.
\label{fig:hubble}}
\end{figure}

Fig. \ref{fig:hubble} indicates the behavior of the Hubble parameter normalized to the Hubble constant $H/H_0$ as a function of $a$. The parameters are the same as these of Fig. \ref{fig:phi}.

\begin{figure}[h!]
\includegraphics[width=90mm]{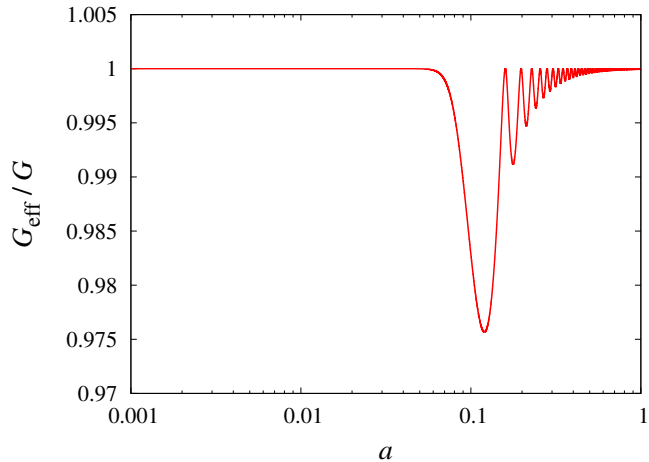}
\caption{Oscillation of the effective gravitational constant normalized to Newton's gravitational constant $G_{\rm eff}/G$. The abscissa is the scale factor $a$.
The parameters are the same as these of Fig. \ref{fig:phi}.
\label{fig:ggg}}
\end{figure}

Fig. \ref{fig:ggg} shows the effective gravitational constant $G_{\rm eff}$ given by \cite{mor1990} \cite{mor1991} \footnote{The scalar field  $\psi$ used in the present work corresponds to $\phi$ in \citet{mor1990,mor1991}. Hence, the quantity $\phi$ used in the present study is equal to $\sqrt{4\pi G/3c^4}$ times Morikawa's $\phi$. See also Eq. (\ref{eq:psiphi}) of the present work.}
\begin{equation}
\frac{G_{\rm eff}}{G}=\frac{1}{1-6\xi\phi^2},  \label{EqGeff}
\end{equation}
where $G$ is Newton's gravitational constant. The abscissa is the scale factor $a$.

In the early epoch, the scalar field starts from a stationary state. Then, it begins to oscillate (Fig. \ref{fig:phi}). The Hubble parameter and the effective gravitational constant also begin to oscillate almost simultaneously (Fig. \ref{fig:hubble}, Fig. \ref{fig:ggg}). The amplitude of the oscillation of the scalar field has been damped until the present time (Fig. \ref{fig:phi}). 

\section{OBSERVATIONAL TESTS \label{obstest}}

In this section, we study the cosmological constraints on the scalar field model of Hirano et al. (2008) \cite{hir2008a} obtained from SN Ia \cite{hic2009}, CMB anisotropies \cite{kom2009}, and BAO \cite{eis2005} observations. Finally, for this model, we investigate variation of the effective gravitational constant in order to test the validity of the model.

In scalar field models \cite{hir2008a}, the scale of the periodic redshift spacings is determined by the mass parameter $m_\phi$ of the scalar field. We fix the mass of the scalar field as $m_\phi=3.2\times 10^{-31}h~{\rm eV}$, such that the scalar field model can account for the redshift spacings of the $127 h^{-1}~{\rm Mpc}$ period \cite{har2008}. Because we want to know the value of the basic parameters of the scalar field model, we fix the constant parameter $q$ introduced by Hirano et al (2008) \cite{hir2008a} as $q=0$. The case of $q=0$ coincides with the model by Morikawa (1990, 1991) \cite{mor1990, mor1991}, Kashino \& Kawabata (1994) \cite{kas1994}, Fukuyama et al. (1997) \cite{fuk1997}, and Hirano et al. (2006) \cite{hir2006}. We fix the Hubble constant as $H_0=72$ ${\rm km~s^{-1} Mpc^{-1}}$ \cite{fre2001}. 

For the other parameters of the scalar field model \cite{hir2008a}, we carry out a detailed investigation of the allowed parameter region using the following observational data.

\subsection{Observational Data}

\subsubsection{Type Ia Supernovae}

We first use the SN Ia constitution dataset, which includes 397 SN Ia \cite{hic2009}. The 90 SN Ia from the CfA3 sample with low redshifts are added to the 307 SN Ia union sample \cite{kow2008}.
 
The dataset gives the distance modulus at redshift $\mu_{obs}(z_i)$.

The theoretical distance modulus (for a given model) is defined by
\begin{equation}
\mu(z)=5\log_{10}{D_L}+\mu_0,
\end{equation}
where $D_L$ is the Hubble free luminosity distance given by
\begin{equation}
D_L=(1+z)\int^z_0\frac{H_0}{H(z^{\prime})}dz^{\prime},
\end{equation}
and $\mu_0$ is
\begin{equation}
\mu_0=5\log_{10}{\left(\frac{{H_0}^{-1}}{Mpc}\right)} + 25 = 42.38 - 5\log_{10}{h}, \label{eq:mu}
\end{equation}
where $h$ is the Hubble constant $H_0$ in units of $100~{\rm km~s^{-1}~Mpc^{-1}}$.

We minimize the statistical $\chi^2$ function (which determines the likelihood function of the parameters) of the model parameters. For the SN Ia data we have
\begin{equation}
\chi_{\rm SN Ia}^2=\sum^{397}_{i=1}\left[\frac{\mu(z_i) - \mu_{obs}(z_i)}{\sigma_{\mu}(z_i)}\right]^2,
\end{equation}
where $\sigma_{\mu}$ is the total uncertainty of the distance modulus \cite{hic2009}.

Because the nuisance parameter $\mu_0$ (Eq. (\ref{eq:mu})) is model-independent, we analytically marginalize it as follows:
\begin{equation}
\chi_{\rm SN Ia}^2=a-\frac{b^2}{c},
\end{equation}
where
\begin{equation}
a=\sum^{397}_{i=1}\frac{[\mu(z_i) - \mu_{obs}(z_i)]^2}{\sigma_{\mu}^2(z_i)},
\end{equation}
\begin{equation}
b=\sum^{397}_{i=1}\frac{\mu(z_i) - \mu_{obs}(z_i)}{\sigma_{\mu}^2(z_i)},
\end{equation}
and
\begin{equation}
c=\sum^{397}_{i=1}\frac{1}{\sigma_{\mu}^2(z_i)}.
\end{equation}
We use the $\chi_{\rm SN Ia}^2$ function in combination with that of the CMB and BAO data.

\subsubsection{Cosmic Microwave Background}

The CMB shift parameter is one of the least model-dependent parameters extracted
from the CMB data. Because this parameter involves a large redshift behavior ($z\sim 1000$), it gives a complementary bound to the SN Ia data ($z \hspace{0.3em}\raisebox{0.4ex}{$<$}\hspace{-0.75em}\raisebox{-.7ex}{$\sim$}\hspace{0.3em} 2$). The shift parameter $R$ is defined as
\begin{equation}
R=\sqrt{\Omega_{m,0}}\int_{0}^{z_{CMB}}{\frac{H_0}{H(z)}dz},
\end{equation}
where $z_{CMB}$ is the redshift at recombination. Although $z_{CMB}$ depends on the matter density $\Omega_{m,0}$ and on the baryon density $\Omega_{b,0}$ at the $\sim$ 2 percent level, we fix this redshift to the 5-year WMAP maximum-likelihood value of $z_{CMB}$ = 1090 \cite{kom2009}.

Using the 5-year WMAP data of $R_{obs} = 1.710 \pm 0.019$ \cite{kom2009}, the $\chi^2$ function for CMB is
\begin{equation}
\chi_{\rm CMB}^2=\left[\frac{R - 1.710}{0.019}\right]^2.
\end{equation} 

\subsubsection{Baryon Acoustic Oscillation}

Observations of large-scale galaxy clustering provide the signatures of the BAO. We use the measurement of the BAO peak in the distribution of luminous red galaxies (LRGs) observed in the Sloan Digital Sky Survey (SDSS) \cite{eis2005}, which gives
\begin{equation}
A_{obs}=0.469\times\left(\frac{0.96}{0.98}\right)^{-0.35}\pm 0.017,
\end{equation}
where the parameter $A$ is calculated by
\begin{equation}
A=\sqrt{\Omega_{m,0}}\left(\frac{H_0}{H(z_{BAO})}\right)^{1/3}\left[\frac{1}{z_{BAO}}\int_0^{z_{BAO}}\frac{H_0}{H(z)}dz\right]^{2/3},
\end{equation}
and $z_{BAO} = 0.35$. The $\chi^2$ function for BAO is
\begin{equation}
\chi_{\rm BAO}^2=\left[\frac{A - 0.469\times\{(0.96/0.98)^{-0.35}\}}{0.017}\right]^2.
\end{equation}

To determine the best value and the allowed region of the parameters, we use the maximum likelihood method and minimize the following quantity:
\begin{equation}
\chi^2=\chi_{\rm SN Ia}^2+\chi_{\rm CMB}^2+\chi_{\rm BAO}^2. \label{chi2}
\end{equation}

\subsection{Numerical Results}

We now present our main results for the constraints from the observational data described in the previous subsection.

\begin{figure}[h!]
\includegraphics[width=90mm]{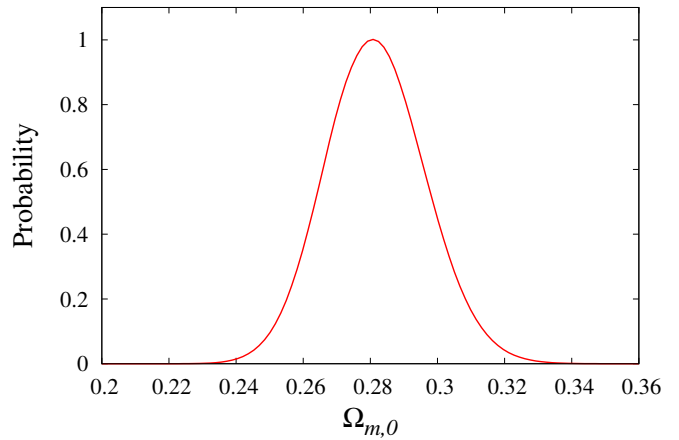}
\caption{1D probability distribution of the energy density parameter of matter $\Omega_{m,0}$ for the oscillating scalar field model from the combination of the SN Ia, CMB, and BAO data.
\label{fig:d1omegam}}
\end{figure}

In Fig. \ref{fig:d1omegam}, we plot the probability distribution of the energy density parameter of matter $\Omega_{m,0}$ for the oscillating scalar field model \cite{hir2008a} from the combination of the SN Ia, CMB, and BAO data, where the parameters $\Omega_{\phi,0}$ and $\xi$ are marginalized.

The best fit value is $\Omega_{m,0}=0.28$. This is similar to the value indicated in the $\Lambda$CDM model. We obtained the comparatively stringent constraint as follows.
\begin{equation}
\Omega_{m,0}=0.28\pm 0.03~~~~~~~~(95\%~{\rm C.L.})
\end{equation}

\begin{figure}[h!]
\includegraphics[width=90mm]{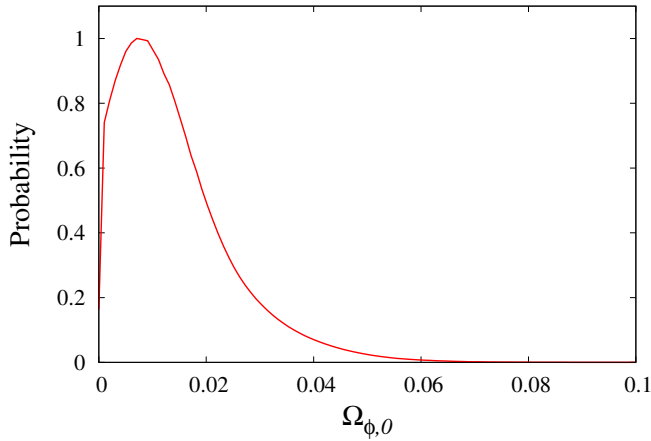}
\caption{1D probability distribution of the energy density parameter of the scalar field $\Omega_{\phi,0}$ for the oscillating scalar field model from the combination of SN Ia, CMB, and BAO data.
\label{fig:d1omegaphi}}
\end{figure}

Fig. \ref{fig:d1omegaphi} shows the probability distribution of the energy density parameter of the scalar field $\Omega_{\phi,0}$ for the oscillating scalar field model \cite{hir2008a} from the combination of SN Ia, CMB, and BAO data, where $\Omega_{m,0}$ and $\xi$ are marginalized.

The best fit value of the energy density parameter of the scalar field $\Omega_{\phi,0}= 0.01$, but not $0$.

We obtained the very stringent constraint as follows.
\begin{equation}
\Omega_{\phi,0} < 0.035~~~~~~~~(95\%~{\rm C.L.})
\end{equation}

We find that the energy density parameter of the scalar field $\Omega_{\phi,0}$ must be small in the universe at present.

\begin{figure}[h!]
\includegraphics[width=90mm]{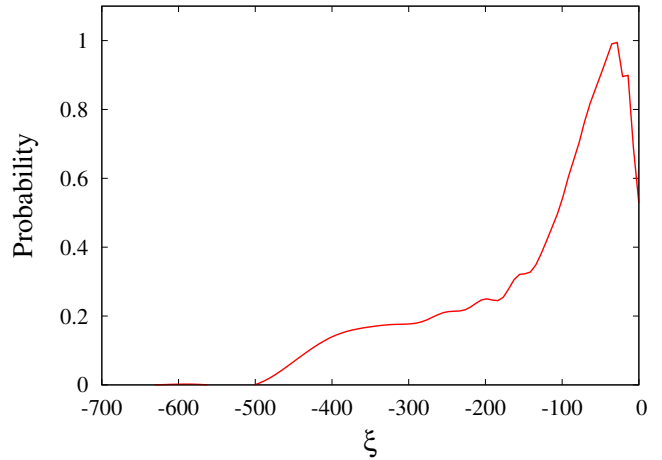}
\caption{1D probability distribution of the coupling constant $\xi$ for the oscillating scalar field model from the combination of the SN Ia, CMB, and BAO data.
\label{fig:d1xi}}
\end{figure}

Fig. \ref{fig:d1xi} shows the probability distribution of the coupling constant $\xi$ for the oscillating scalar field model \cite{hir2008a} from the combination of the SN Ia, CMB, and BAO data, where $\Omega_{m,0}$ and $\Omega_{\phi,0}$ are marginalized. Note that Fukuyama et al. (1997) \cite{fuk1997} pointed out that the use of negative values for the coupling constant $\xi$ yields a succession of mini-inflationary (concave shaped) cosmic expansions with time, thereby prolonging the cosmological age. We therefore employ negative values for $\xi$. We search by step of logarithm for parameter $\xi$.  

The best fit value of the coupling constant is $\xi= -25$.
The constraint is
\begin{equation}
\xi > -158~~~~~~~~(95\%~{\rm C.L.})~~~~~~~~~~~({\rm in~the~range~} \xi \le 0)
\end{equation}

The absolute value of $\xi$ cannot be very large.

\begin{figure}[h!]
\includegraphics[width=90mm]{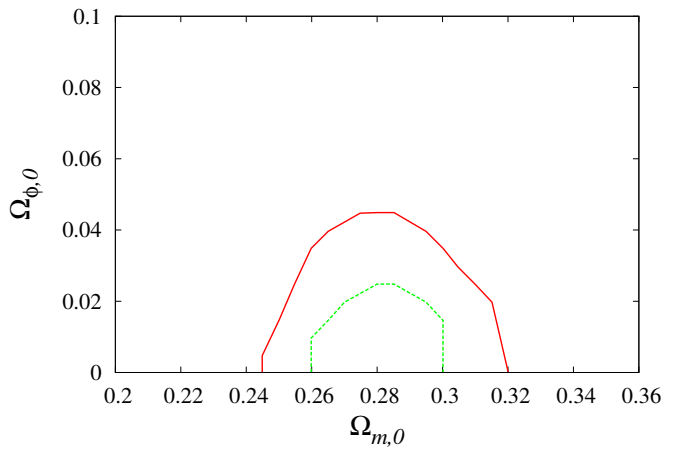}
\caption{2D probability contours in the ($\Omega_{m,0}$, $\Omega_{\phi,0}$)-plane for the oscillating scalar field model. The dotted (green) and solid (red) contours show the $1\sigma$ (68\%) and $2\sigma$ (95\%) confidence limits, respectively, from a combined analysis of the SN Ia, CMB, and BAO data.
\label{fig:d2_m_phi}}
\end{figure}

\begin{figure}[h!]
\includegraphics[width=90mm]{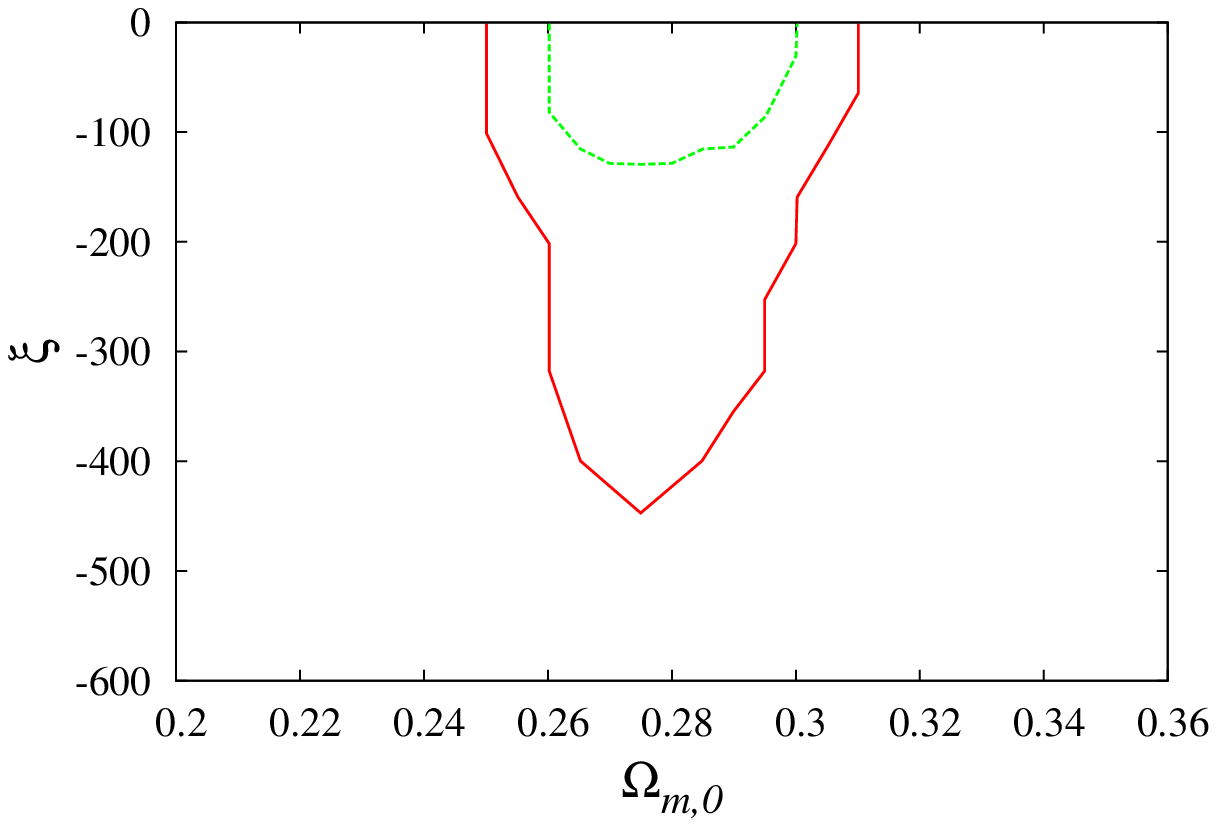}
\caption{2D probability contours in the ($\Omega_{m,0}$, $\xi$)-plane for the oscillating scalar field model. The dotted (green) and solid (red) contours show the $1\sigma$ (68\%) and $2\sigma$ (95\%) confidence limits, respectively, from a combined analysis of the SN Ia, CMB, and BAO data.
\label{fig:d2_m_xi}}
\end{figure}

\begin{figure}[h!]
\includegraphics[width=90mm]{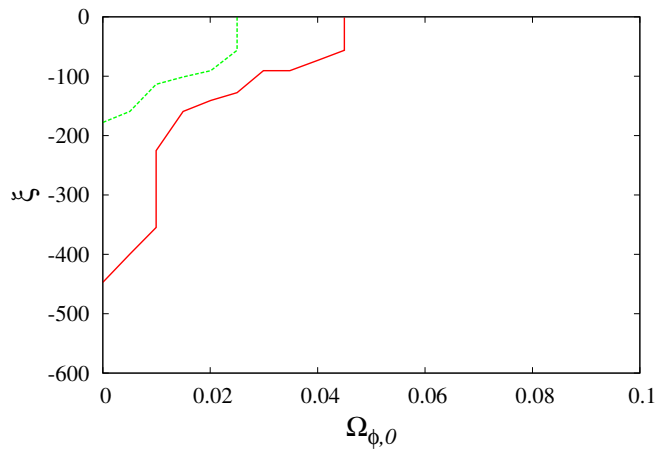}
\caption{2D probability contours in the ($\Omega_{\phi,0}$, $\xi$)-plane for the oscillating scalar field model. The dotted (green) and solid (red) contours show the $1\sigma$ (68\%) and $2\sigma$ (95\%) confidence limits, respectively, from a combined analysis of the SN Ia, CMB, and BAO data.
\label{fig:d2_phi_xi}}
\end{figure}

Fig. \ref{fig:d2_m_phi}, Fig. \ref{fig:d2_m_xi}, and Fig. \ref{fig:d2_phi_xi} show the 2D probability contours for the oscillating scalar field model \cite{hir2008a}. The dotted (green) and solid (red) contours show the $1\sigma$ (68\%) and $2\sigma$ (95\%) confidence limits, respectively, from a combined analysis of the SN Ia, CMB, and BAO data. The other parameters are marginalized. 

The best fit value of $\Omega_{m,0}$ is $0.28$, and the best fit value of $\Omega_{\phi,0}$ is close to but not equal to $0$. The absolute value of $\xi$ cannot be very large.

The amplitude of $N$ from $z$ data (the picket-fence structure of the $N$--$z$ relation) mainly depends on $\Omega_{\phi,0}$ and $\xi$ in the oscillating scalar field model. The observational amplitude of the galaxy number count has a high uncertainty, but in the scalar field model, since the most likely value of $\Omega_{\phi,0}$ is small, the amplitude cannot be large. 

However, the scale of the periodic redshift spacings is determined by the mass parameter $m_\phi$ of the scalar field. The scalar field model can account for the redshift spacings for the $127 h^{-1}~{\rm Mpc}$ period \cite{har2008}, assuming a mass of the scalar field of $m_\phi=3.2\times 10^{-31}h~{\rm eV}$. Because the most likely values of $\Omega_{\phi,0}$ and $\xi$ are not $0$, the scalar field model has the possibility of accounting for the periodic structure in the $N$--$z$ relation of galaxies. \\

\begin{figure}[h!]
\includegraphics[width=90mm]{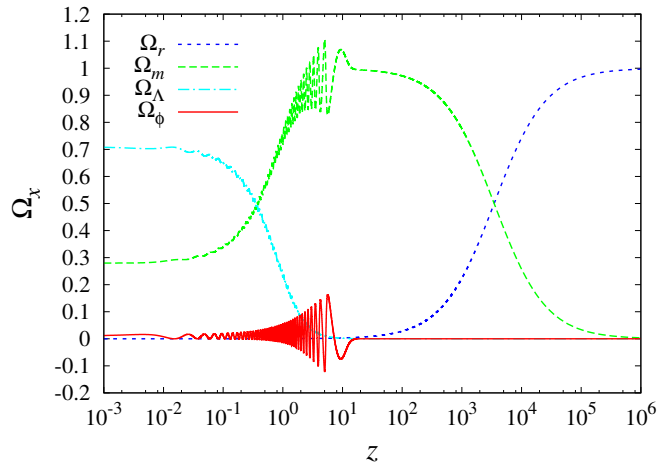}
\caption{Normalized energy density of radiation $\Omega_r$, matter $\Omega_m$, cosmological constant $\Omega_{\Lambda}$, and scalar field $\Omega_{\phi}$, versus the redshift $z$ in the oscillating scalar field model \cite{hir2008a} with the best fit parameters: $\Omega_{m,0}=0.28$, $\Omega_{\phi,0}=0.01$, $\xi=-25$.
\label{fig:omega}}
\end{figure}

Fig. \ref{fig:omega} shows the normalized energy density of radiation $\Omega_r$, matter $\Omega_m$, cosmological constant $\Omega_{\Lambda}$, and scalar field $\Omega_{\phi}$ versus the redshift $z$ in the oscillating scalar field model \cite{hir2008a} with the best fit parameters: $\Omega_{m,0}=0.28$, $\Omega_{\phi,0}=0.01$, $\xi=-25$, where $\Omega_{\phi} = (8\pi G/3H^2)\rho_{\phi}$. $\rho_{\phi}$ is the energy density of scalar field defined by Eq. (\ref{rhopsi}). We find that the normalized energy density of scalar field $\Omega_{\phi}$ is comparatively small in all era.

In Table \ref{table1}, we list the best fit parameters, the $\chi^2$ values (Eq. (\ref{chi2})), and the differences of the Akaike information criterion (AIC) \cite{aka1974} and the Bayesian information criterion (BIC) \cite{sch1978}, for the scalar field model and the $\Lambda$CDM model, from a combined analysis of the SN Ia, CMB, and BAO data. The definitions of AIC and BIC are
\begin{equation}
{\rm AIC} = -2\ln{L} + 2k,
\end{equation}
\begin{equation}
{\rm BIC} = -2\ln{L} + k\ln{N},
\end{equation}
where $L$ is the maximum likelihood, $k$ is the number of free model parameters, and $N$ is the number of data points used in the fit. The $\chi^2$ value for the scalar field model is smaller than that for the $\Lambda$CDM model. However, the values of AIC and BIC for the $\Lambda$CDM model are smaller than that for the scalar field model, because the number of free parameters of the scalar field model is two more than that of the $\Lambda$CDM model.

\begin{table*}
\caption{Results of observational tests from the combination of SN Ia, CMB, and BAO data. \label{table1}}
\begin{tabular}{l l l r r}
\hline\hline
Model  &  Best fit parameters  &  $\chi^2$  &  $\Delta$AIC  &  $\Delta$BIC \\
\hline
$\Lambda$CDM  &  \footnote{We fix the other parameters at $H_0=72$ ${\rm km~s^{-1} Mpc^{-1}}$.}$\Omega_{m,0}=0.28$  &  466.73  &  0.00  &  0.00 \\ 
Oscillating Scalar Field \cite{hir2008a}~~~~~  &  \footnote{We fix the other parameters at $H_0=72$ ${\rm km~s^{-1} Mpc^{-1}}$, $m_\phi=3.2\times 10^{-31}h~{\rm eV}$, and $q=0$.}$\Omega_{m,0}=0.28$,~ $\Omega_{\phi,0}=0.01$,~$\xi=-25$ ~~~~~  &  465.86~~~~  &  3.13  &  ~~~~ 11.11 \\
\hline\hline
\end{tabular}
\end{table*}

\subsection{Variation of the Effective Gravitational Constant}

In order to test the validity of the model, we compute the variation of the effective gravitational constant.

From Eq. (\ref{EqGeff}), the variation of the effective gravitational constant is
\begin{equation}
\frac{G^{\hspace{0.5mm}{\displaystyle \prime}}_{\rm eff}}{G_{\rm eff}}=\frac{12\xi\phi\phi^{\prime}}{1-6\xi\phi^2}, 
\end{equation}
where prime designates the derivative with respect to the proper time $t$.

\begin{figure}[h!]
\includegraphics[width=90mm]{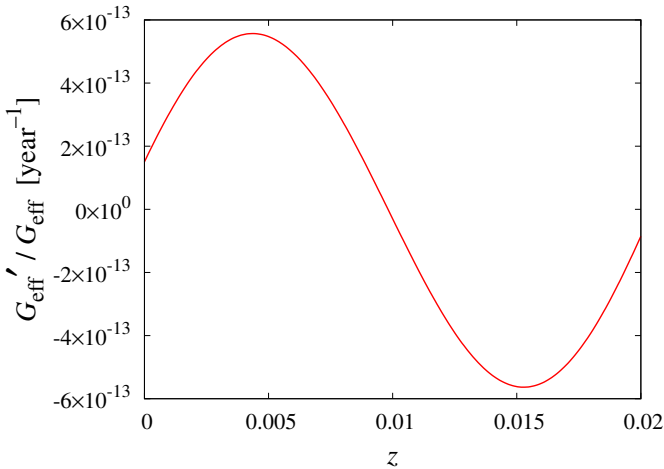}
\caption{Variation of the effective gravitational constant $G^{\hspace{0.5mm}{\displaystyle \prime}}_{\rm eff}/G_{\rm eff}$ versus the redshift $z$ for the oscillating scalar field model with the best fit parameters in Table \ref{table1} (The parameters same as Fig. \ref{fig:omega}).
\label{fig:gvari}}
\end{figure}

In Fig. \ref{fig:gvari}, we plot the variation of the effective gravitational constant $G^{\hspace{0.5mm}{\displaystyle \prime}}_{\rm eff}/G_{\rm eff}$ versus the redshift $z$ for the oscillating scalar field model \cite{hir2008a} with the best fit parameters in Table \ref{table1}. In our model, we get $G^{\hspace{0.5mm}{\displaystyle \prime}}_{\rm eff}/G_{\rm eff}=1.5\times 10^{-13}~{\rm yr^{-1}}$ at present. Because of the small value of $\Omega_{\phi,0}$, the variation of the effective gravitational constant is modest.

Observationally, there are severe constraints on this quantity. From lunar laser ranging tests \cite{wil2004}, the constraint is reported to be 
\begin{equation}
\frac{G^{\hspace{0.5mm}{\displaystyle \prime}}_{\rm eff}}{G_{\rm eff}}= (4 \pm 9) \times 10^{-13}~{\rm yr^{-1}}.
\end{equation}

Although it is known that the temporal change of the gravitational constant affects solar evolution, our prediction is not inconsistent with the bound from the observation.

\section{CONCLUSION \label{conclu}}

According to a Fourier analysis by Hartnett \& Hirano \cite{har2008}, the $N$--$z$ relation indicates that galaxies have preferred periodic redshift spacings. The oscillating scalar field model \cite{mor1990, mor1991, hir2008a} can account for the redshift spacings:  the scalar field oscillates around the potential minimum, inducing an oscillation of the Hubble parameter, due to the curvature coupling of $\phi$. This gives a periodic structure as an apparent or illusionary effect in the $N$--$z$ relation, preserving the cosmological principle. 

Assuming the scalar field model of Hirano et al. (2008) \cite{hir2008a} with $m_\phi=3.2\times 10^{-31}h~{\rm eV}$ (such that the scalar field model can account for the redshift spacings of the $127 h^{-1}~{\rm Mpc}$ period \cite{har2008}) we obtained the following constraints for the model parameters: $\Omega_{m,0}=0.28\pm 0.03$ (95\% C.L.), $\Omega_{\phi,0} < 0.035$ (95\% C.L.), $\xi > -158$ (95\% C.L.) (in the range $\xi \le 0$). We obtained the best fit values $\Omega_{m,0}=0.28$, $\Omega_{\phi,0}= 0.01$, and $\xi= -25$.

The amplitude of $N$ from the $z$ data (the picket-fence structure of the $N$--$z$ relation) depends mainly on $\Omega_{\phi,0}$ and $\xi$ in the oscillating scalar field model. The observational amplitude of $N$ has a high uncertainty, though in the scalar field model, since the best fit value of $\Omega_{\phi,0}$ is small, the amplitude cannot be large. However, 
because the best fit values of $\Omega_{\phi,0}$ and $\xi$ are not $0$, the scalar field model has the possibility of accounting for the periodic structure in the $N$--$z$ relation of galaxies. The variation of the effective gravitational constant in the scalar field model is not inconsistent with the bound from observation.

\section*{ACKNOWLEDGEMENTS}
The authors are grateful to John G. Hartnett for useful comments and discussions.

\bibliography{hirano}

\end{document}